\documentclass[aps,pra,showpacs,superscriptaddress,preprint]{revtex4}
\usepackage{amssymb}
\usepackage{amsmath}
\usepackage{graphicx}

\catcode`ð=\active
 \defð{\u{g}}
 \catcode`Ð=\active
 \defÐ{\u{G}}
 \catcode`Ý=\active
\defÝ{\. I}
 \catcode`ö=\active
\defö{\"{o}}
 \catcode`Ö=\active
 \defÖ{\"O}
 \catcode`ü=\active
 \defü{\"{u}}
 \catcode`Ü=\active
 \defÜ{\"{U}}
 \catcode`Þ=\active
\defÞ{\c{S}}
 \catcode`þ=\active
 \defþ{\c{s}}
 \catcode`ý=\active
 \defý{{\i}}
 \catcode`ç=\active
\defç{\d{c}}
 \catcode`Ç=\active
\defÇ{\d{C}}

\begin{document}

\title{Approximate bound state solutions of Dirac equation with Hulth\'{e}n
potential including Coulomb-like tensor potential }
\author{Sameer M. Ikhdair}
\email[E-mail: ]{sikhdair@neu.edu.tr}
\affiliation{Physics Department, Near East University, Nicosia, North Cyprus, Turkey}
\author{Ramazan Sever}
\email[E-mail: ]{sever@metu.edu.tr}
\affiliation{Physics Department, Middle East Technical University, 06531, Ankara,Turkey}
\date{%
\today%
}

\begin{abstract}
We solve the Dirac equation approximately for the attractive scalar $S(r)$
and repulsive vector $V(r)$ Hulth\'{e}n potentials including a Coulomb-like
tensor potential with arbitrary spin-orbit coupling quantum number $\kappa .$
In the framework of the spin and pseudospin symmetric concept, we obtain the
analytic energy spectrum and the corresponding two-component upper- and
lower-spinors of the two Dirac particles by means of the Nikiforov-Uvarov
method in closed form. The limit of zero tensor coupling and the
non-relativistic solution are obtained. The energy spectrum for various
levels is presented for several $\kappa $ values under the condition of
exact spin symmetry in the presence or absence of tensor coupling.

\end{abstract}

\maketitle

\newpage

\section{Introduction}

Within the framework of the Dirac equation the spin symmetry arises if the
magnitude of the attractive Lorentz scalar potential $S(r)$ and the
time-component repulsive vector potential are nearly equal, $S(r)\sim V(r)$
in nuclei (\textit{i.e.}, when the difference potential $\Delta
(r)=V(r)-S(r)=C_{s}=$ constant$).$ However, the pseudospin symmetry occurs
when $S(r)\sim -V(r)$ are nearly equal (\textit{i.e.}, when the sum
potential $\Sigma (r)=V(r)+S(r)=C_{ps}=$ constant$)$ [1-3]$.$ The bound
states of nucleons seem to be sensitive to some mixtures of these
potentials. The cases $\Delta (r)=0$ and $\Sigma (r)=0$ correspond to $SU(2)$
symmetries of the Dirac Hamiltonian [3]. The spin symmetry is relevant for
mesons [4]. The pseudospin symmetry concept has been applied to many systems
in nuclear physics and related areas [2-7]. Further, it is used to explain
features of deformed nuclei [8], the super-deformation [9] and to establish
an effective nuclear shell-model scheme [5,6,10]. The pseudospin symmetry
appeared in nuclear physics refers to a quasi-degeneracy of the
single-nucleon doublets and can be characterized with the non-relativistic
quantum numbers $(n,l,j=l+1/2)$ and $(n-1,l+2,j=l+3/2),$ where $n,$ $l$ and $%
j$ are the single-nucleon radial, orbital and total angular momentum quantum
numbers for a single particle, respectively [5,6]. The total angular
momentum is given as $j=\widetilde{l}+\widetilde{s},$ where $\widetilde{l}%
=l+1$ is a pseudo-angular momentum and $\widetilde{s}=1/2$ is a pseudospin
angular momentum. In real nuclei, the pseudospin symmetry is only an
approximation and the quality of approximation depends on the
pseudo-centrifugal potential and pseudospin orbital potential [11]. The
Dirac Hamiltonian with vector and scalar potentials quadratic in space
coordinates has been studied [12]. It was shown that the the Dirac equation
can be solved exactly for the cases $\Delta (r)=0$ and $\Sigma (r)=0.$ In
addition, the linear tensor potential and mixture of quadratic scalar and
vector potentials are studied [13]. It is shown that a linear tensor
potential with quadratic $\Delta (r)$ or $\Sigma (r)$ generates a
harmonic-oscillator-like second order differential equation which can be
solved analytically. Recently, Akcay [14,15] has shown that the Dirac
equation for scalar and vector quadratic potentials including the
Coulomb-like tensor potential with the spin and pseudospin symmetries can be
solved exactly. These results in Dirac equation with quadratic potential
plus a centrifugal-like potential can also be solved analytically. Tensor
coupling potentials are added as spin-orbit coupling terms to the Dirac
Hamiltonian by making the substitution $\overrightarrow{p}\rightarrow
\overrightarrow{p}-im\omega \mathbf{\beta .}\widehat{r}U(r)$ [16,17]. Tensor
couplings and exactly solvable tensor potential have been used to
investigate nuclear properties [18-20] and have also some physical
applications [21,22].

In the past years, there has been much interest in the solution of the
relativistic Dirac and Klein-Gordon equations [1,12,23-26]. For instance,
some authors have solved these equations for several physical potentials,
such as the Woods-Saxon potential [23-26], the Morse potential [27], the
Hulth\'{e}n potential [28], the Eckart potential [29-31], the P\"{o}%
schl-Teller potential [32,33] and the Scarf-type potential [34], etc.

Recently, many works have been done to solve the Dirac equation so to obtain
the energy equation and the two-component spinor wave functions. Jia \textit{%
et al} [35] employed an improved approximation scheme to deal with the new
centrifugal spin-orbit term $\kappa \left( \kappa +1\right) r^{-2}$ in the
second order differential equation that results from the Dirac equation and
to solve it for the generalized P\"{o}schl-Teller potential for arbitrary
spin-orbit quantum number $\kappa .$ Zhang \textit{et al} [36] solved the
Dirac equation with equal Scarf-type scalar and vector potentials by the
method of the supersymmetric (SUSY) quantum mechanics, shape invariance
approach and by alternative methods. Zou \textit{et al} [37] solved the
Dirac equation with equal Eckart scalar and vector potentials in terms of
SUSY quantum mechanical method, shape invariance approach and function
analysis method. Wei and Dong [38] obtained approximately the analytical
bound state solutions of the Dirac equation with the Manning-Rosen for
arbitrary $\kappa .$ Thylwe [39] presented the approach inspired by
amplitude-phase method in analyzing the radial Dirac equation to calculate
phase shifts by including the spin- and pseudo-spin symmetries of
relativistic spectra. Alhaidari [40] solved Dirac equation by separation of
variables in spherical coordinates for a large class of non-central
electromagnetic potentials. Berkdemir and Sever [41] investigated
systematically the pseudospin symmetric solution of the Dirac equation for
spin $1/2$ particles moving within the Kratzer potential connected with an
angle-dependent potential. Recently, we have also solved the spin and
pseudospin symmetric Dirac equation with arbitrary spin-orbit centrifugal
term for generalized Woods-Saxon potential [42] and Rosen-Morse potential
[43] by means of the Nikiforov-Uvarov (NU) method.

In this paper, it is worth to investigate the solution of the Dirac equation
for scalar and vector Hulth\'{e}n potential for $\Delta (r)$ or $\Sigma (r)$
together with Coulomb shape tensor coupling potential which can be solved
analytically by using an improved approximation scheme introduced in Refs.
[44,45] to deal with the resulting centrifugal and pseudo-centrifugal terms $%
\kappa \left( \kappa \pm 1\right) r^{-2}$. The Coulomb-like tensor potential
preserves the form of the Hulth\'{e}n potential but generates a new
spin-orbit centrifugal terms $\Lambda \left( \Lambda \pm 1\right) r^{-2},$
where $\Lambda $ is a new spin-orbit quantum number. This provides a
possibility for generating a different form of spin-orbit coupling which
might have some physical applications.

The Hulth\'{e}n potential, widely used for the description of the
nucleon-heavy nucleus interactions, takes the following form (see [44] and
the references therein):%
\begin{equation}
V_{H}(r)=-\frac{V_{0}}{e^{r/r_{0}}-1},\text{ }r_{0}=\delta ^{-1},\text{ }%
V_{0}=Ze^{2}\delta ,
\end{equation}%
where $V_{0}$ is the potential depth, $\delta $ is the screening range
parameter and $r_{0}$ represents the spatial range. If the potential is used
for atoms, then $V_{0}=Z\delta $ (in the relativistic units $\hbar =c=e=1$),
where $Z$ is identified as the atomic number. The Hulth\'{e}n potential
behaves like the Coulomb potential near the origin $($i.e., $r\rightarrow 0$
or $r\ll r_{0})$ $V_{C}(r)=-Ze^{2}/r$ , but decreases exponentially in the
asymptotic region when $r\gg 0,$ so its capacity for bound states is smaller
than the Coulomb potential. This potential has been applied to a number of
areas such as nuclear and particle physics [46-48], atomic physics [49,50],
molecular physics [51,52] and chemical physics [53], etc.

In the presence of the spin and pseudospin symmetry, we investigate the
bound state energy eigenvalues and corresponding upper and lower spinor wave
functions for arbitrary spin-orbit $\kappa $ quantum number in the framework
of the NU method [54-56]. We also show that the spin and pseudospin
symmetric Dirac solutions can be reduced to the $S(r)=V(r)$ and $S(r)=-V(r)$
in the cases of exact spin symmetry limitation $\Delta (r)=0$ and pseudospin
symmetry limitation $\Sigma (r)=0,$ respectively. Furthermore, the solutions
obtained for the Dirac equation can be easily reduced to the Schr\"{o}dinger
solutions when a parametric transformation is applied.\

In what follows, we first review the NU method and present a parametric
generalization in Sect. 2. In the presence of spin and pseudo-spin
symmetries, we obtain the bound state solutions of the Dirac equation with
scalar and vector Hulth\'{e}n potentials including the Coulomb-like tensor
interaction, the limit of zero tensor coupling and the non-relativistic
limits by applying a suitable transformation in Sect. 3. The relevant
concluding remarks are given in Sect. 4.

\section{The Nikiforov-Uvarov Method}

The NU method [54] is briefly outlined here. It is based on solving the
second-order differential equation of hypergeometric-type by means of
special orthogonal functions:
\begin{equation}
\psi ^{\prime \prime }(r)+\frac{\widetilde{\tau }(r)}{\sigma (r)}\psi
^{\prime }(r)+\frac{\widetilde{\sigma }(r)}{\sigma ^{2}(r)}\psi (r)=0,
\end{equation}%
where $\sigma (r)$ and $\widetilde{\sigma }(r)$ are polynomials at most of
second-degree, $\widetilde{\tau }(r)$ is a first-degree polynomial and $\psi
(r)$ is function of the hypergeometric type. In order to find a particular
solution for Eq. (2), we choose $\psi (r)$ as follows:%
\begin{equation}
\psi (r)=\phi (r)y(r),
\end{equation}%
which leads to a new second-order hypergeometric-type equation of the form%
\begin{equation}
y^{\prime \prime }(r)+A(r)y^{\prime }(r)+B(r)y(r)=0,
\end{equation}%
with $A(r)$ and $B(r)$ are taken to be $\tau (r)/\sigma (r)$ and $\widetilde{%
\sigma }(r)/\sigma ^{2}(r),$ respectively, where $\widetilde{\sigma }%
(r)=\lambda \sigma (r),$ $\lambda $ is a constant and $\tau (r)$ is a
polynomial of degree at most one$.$ If we impose that $\phi (r)$ satisfies
the following logarithmic equation%
\begin{equation}
\frac{\phi ^{\prime }(r)}{\phi (r)}=\frac{\pi (r)}{\sigma (r)},
\end{equation}%
then we obtain%
\begin{equation}
\tau (r)=\widetilde{\tau }(r)+2\pi (r),\text{ }\tau ^{\prime }(r)<0,
\end{equation}%
where $\pi (r)$ is a polynomial of order at most one. Making use of $\ \phi
^{\prime \prime }(r)/\phi (r)=\left( \phi ^{\prime }(r)/\phi (r)\right)
^{\prime }+\left( \phi ^{\prime }(r)/\phi (r)\right) ^{2},$ we can reduce
Eq. (4) into another hypergeometric-type:%
\begin{equation}
\sigma (r)y^{\prime \prime }(r)+\tau (r)y^{\prime }(r)+\lambda y(r)=0,
\end{equation}%
and the quadratic equation for the polynomial $\pi (r),$%
\begin{equation}
\pi ^{2}(r)+\left[ \widetilde{\tau }(r)-\sigma ^{\prime }(r)\right] \pi (r)+%
\widetilde{\sigma }(r)-k\sigma (r)=0,
\end{equation}%
where%
\begin{equation}
k=\lambda -\pi ^{\prime }(r).
\end{equation}%
The solution of the above quadratic equation for $\pi (r)$ yields%
\begin{equation}
\pi (r)=\frac{1}{2}\left[ \sigma ^{\prime }(r)-\widetilde{\tau }(r)\right]
\pm \sqrt{\frac{1}{4}\left[ \sigma ^{\prime }(r)-\widetilde{\tau }(r)\right]
^{2}-\widetilde{\sigma }(r)+k\sigma (r)},
\end{equation}%
and the weight function can be obtained via
\begin{equation}
\left[ \sigma (r)\rho (r)\right] ^{\prime }=\tau (r)\rho (r),
\end{equation}%
where the prime denotes the differentiation with respect to $r.$ The
expression under the square root sign in Eq. (10) can be arranged as a
polynomial of second order where its discriminant is zero. Hence, an
equation for $k$ is being obtained. After solving such an equation, the $k$
values are determined through the NU method. One, however, is looking for a
family of solutions corresponding to%
\begin{equation}
\lambda =\lambda _{n}=-n\tau ^{\prime }(r)-\frac{1}{2}n\left( n-1\right)
\sigma ^{\prime \prime }(r),\ \ \ n=0,1,2,\cdots .
\end{equation}%
The $y(r)=y_{n}(r)$ which is a polynomial of degree $n$ can be expressed in
terms of the Rodrigues relation:%
\begin{equation}
y_{n}(r)=\frac{B_{n}}{\rho (r)}\frac{d^{n}}{dr^{n}}\left[ \sigma ^{n}(r)\rho
(r)\right] ,
\end{equation}%
where $B_{n}$ is the normalization constant and the weight function $\rho
(r) $ in (13) is the solution of the differential equation (11).

In addition, an eigenvalue solution through the NU method can be set up from
the relationship between $\lambda $ and $\lambda _{n}$ by means of Eqs. (9)
and (12).

For a more simple application of the method, we develop a parametric
generalization of the NU method valid for any central and non-central
exponential-type potential by making change of the independent variables $%
z=z(r).$ Thus, we obtain another generalized hypergeometric equation
\begin{equation}
\left[ z\left( 1-c_{3}z\right) \right] ^{2}w^{\prime \prime }(z)+\left[
z\left( 1-c_{3}z\right) \left( c_{1}-c_{2}z\right) \right] w^{\prime
}(z)+\left( -q_{2}z^{2}+q_{1}z-q_{0}\right) w(z)=0,
\end{equation}%
when compared with Eq. (1) yields%
\begin{equation}
\widetilde{\tau }(z)=c_{1}-c_{2}z,\text{ }\sigma (z)=z\left( 1-c_{3}z\right)
,\text{ }\widetilde{\sigma }(z)=-q_{2}z^{2}+q_{1}z-q_{0},
\end{equation}%
where the parameters $c_{j}$ and $q_{j}$ ($j=0,1,2$) are to be determined
during the solution procedure. Thus, by following the method, we can also
obtain all the analytic polynomials and their relevant constants necessary
for the solution of a radial wave equation. Hence, these analytical
expressions are displayed in Appendix A.

\section{Dirac equation with a tensor coupling}

In spherical coordinates, the Dirac equation for fermionic massive spin-$1/2$
particles moving in attractive scalar $S(r),$ repulsive vector $V(r)$ and a
tensor $U(r)$ potentials reads as (in relativistic units $\hbar =c=1$)
[13,57]:
\begin{equation}
\left[ \mathbf{\alpha }\cdot \mathbf{p+\beta }\left( M+S(r)\right) +V(r)-i%
\mathbf{\beta \mathbf{\alpha }\cdot }\widehat{r}U(r)-E\right] \psi _{n\kappa
}(\mathbf{r})=0,\text{ }\psi _{n\kappa }(\mathbf{r})=\psi (r,\theta ,\phi ),
\end{equation}%
where $E$ is the relativistic energy of the system, $M$ is the mass of the
fermionic particle, $\mathbf{p}=-i\mathbf{\nabla }$ is the three-dimensional
momentum operator, and $\mathbf{\alpha }$ and $\mathbf{\beta }$ are $4\times
4$ Dirac matrices, which have the following forms, respectively,
\begin{equation}
\mathbf{\alpha =}\left(
\begin{array}{cc}
0 & \mathbf{\sigma } \\
\mathbf{\sigma } & 0%
\end{array}%
\right) ,\text{ }\mathbf{\beta =}\left(
\begin{array}{cc}
\mathbf{I} & 0 \\
0 & -\mathbf{I}%
\end{array}%
\right) ,
\end{equation}%
where $\mathbf{I}$ denotes the $2\times 2$ identity matrix and $\mathbf{%
\sigma }$ are three-vector Pauli spin matrices%
\begin{equation}
\sigma _{1}\mathbf{=}\left(
\begin{array}{cc}
0 & 1 \\
1 & 0%
\end{array}%
\right) ,\text{ }\sigma _{2}\mathbf{=}\left(
\begin{array}{cc}
0 & -i \\
i & 0%
\end{array}%
\right) ,\text{ }\sigma _{3}\mathbf{=}\left(
\begin{array}{cc}
1 & 0 \\
0 & -1%
\end{array}%
\right) .
\end{equation}%
For a particle in a spherical (central) field, the total angular momentum
operator $\mathbf{J}$ and the spin-orbit matrix operator $\widehat{\mathbf{K}%
}=-\mathbf{\beta }\left( \mathbf{\sigma }\cdot \mathbf{L}+\mathbf{I}\right) $
commute with the Dirac Hamiltonian, where $\mathbf{L}$ is the orbital
angular momentum operator. For a given total angular momentum $j,$ the
eigenvalues of $\widehat{\mathbf{K}}$ are $\kappa =-(j+1/2)$ for aligned
spin ($s_{1/2},$ $p_{3/2},$ \textit{etc.}) and $\kappa =j+1/2$ for unaligned
spin ($p_{1/2},$ $d_{3/2},$ \textit{etc.}). The spinor wavefunctions can be
classified according to the radial quantum number $n$ and the spin-orbit
quantum number $\kappa $ and can be written using the Pauli-Dirac
representation:%
\begin{equation}
\psi _{n\kappa }(r,\theta ,\phi )=\frac{1}{r}\left(
\begin{array}{c}
F_{n\kappa }(r)Y_{jm}^{l}(\theta ,\phi ) \\
iG_{n\kappa }(r)Y_{jm}^{\widetilde{l}}(\theta ,\phi )%
\end{array}%
\right) ,
\end{equation}%
where $F_{n\kappa }(r)$ and $G_{n\kappa }(r)$ are the radial wave functions
of the upper- and lower-spinor components, respectively, $Y_{jm}^{l}(\theta
,\phi )$ and $Y_{jm}^{\widetilde{l}}(\theta ,\phi )$ are the spherical
harmonic functions coupled to the total angular momentum $j$ and it's
projection $m$ on the $z$ axis. The orbital and pseudo-orbital angular
momentum quantum numbers for spin symmetry $l$ and pseudospin symmetry $%
\widetilde{l}$ refer to the upper- and lower-components, respectively. For a
given spin-orbit quantum number $\kappa =\pm 1,\pm 2,\cdots ,$ the orbital
angular momentum and pseudo-orbital angular momentum are given by $%
l=\left\vert \kappa +1/2\right\vert -1/2$ and $\widetilde{l}=\left\vert
\kappa -1/2\right\vert -1/2,$ respectively$.$ The quasi-degenerate doublet
structure can be expressed in terms of a pseudo-spin angular momentum $%
\widetilde{s}=1/2$ and pseudo-orbital angular momentum $\widetilde{l}$ which
is defined as $\widetilde{l}$ $=l+1$ for aligned spin $j=\widetilde{l}-1/2$
and $\widetilde{l}$ $=l-1$ for unaligned spin $\ j=\widetilde{l}+1/2.$ For
example, ($3s_{1/2},2d_{3/2}$) and ($3\widetilde{p}_{1/2},2\widetilde{p}%
_{3/2}$) can be considered as pseudospin doublets.

Substituting Eq. (19) into Eq. (16) and using the following relations [57]
\begin{subequations}
\begin{equation}
\left( \mathbf{\sigma }\cdot \mathbf{A}\right) \left( \mathbf{\sigma }\cdot
\mathbf{B}\right) =\mathbf{A}\cdot \mathbf{B}+i\mathbf{\sigma }\cdot \left(
\mathbf{A}\times \mathbf{B}\right) ,
\end{equation}%
\begin{equation}
\left( \mathbf{\sigma }\cdot \mathbf{P}\right) =\mathbf{\sigma }\cdot
\widehat{\mathbf{r}}\left( \widehat{\mathbf{r}}\cdot \mathbf{P+i}\frac{%
\mathbf{\sigma }\cdot \mathbf{L}}{r}\right) ,
\end{equation}%
and properties
\end{subequations}
\begin{subequations}
\begin{equation}
\left( \mathbf{\sigma }\cdot \mathbf{L}\right) Y_{jm}^{\widetilde{l}}(\theta
,\phi )=\left( \kappa -1\right) Y_{jm}^{\widetilde{l}}(\theta ,\phi ),
\end{equation}%
\begin{equation}
\left( \mathbf{\sigma }\cdot \mathbf{L}\right) Y_{jm}^{l}(\theta ,\phi
)=-\left( \kappa +1\right) Y_{jm}^{l}(\theta ,\phi ),
\end{equation}%
\begin{equation}
\left( \mathbf{\sigma }\cdot \widehat{r}\right) Y_{jm}^{\widetilde{l}%
}(\theta ,\phi )=-Y_{jm}^{l}(\theta ,\phi ),
\end{equation}%
\begin{equation}
\left( \mathbf{\sigma }\cdot \widehat{r}\right) Y_{jm}^{l}(\theta ,\phi
)=-Y_{jm}^{\widetilde{l}}(\theta ,\phi ),
\end{equation}%
we obtain the following two radial coupled Dirac equations for the spinor
components:
\end{subequations}
\begin{subequations}
\begin{equation}
\left( \frac{d}{dr}+\frac{\kappa }{r}-U(r)\right) F_{n\kappa }(r)=\left(
M^{2}+E_{n\kappa }-\Delta (r)\right) G_{n\kappa }(r),
\end{equation}%
\begin{equation}
\left( \frac{d}{dr}-\frac{\kappa }{r}+U(r)\right) G_{n\kappa }(r)=\left(
M^{2}-E_{n\kappa }+\Sigma (r)\right) F_{n\kappa }(r),
\end{equation}%
where $\Delta (r)=V(r)-S(r)$ and $\Sigma (r)=V(r)+S(r)$ are the difference
and sum potentials, respectively. By eliminating $G_{n\kappa }(r)$ in Eq.
(22a) and $F_{n\kappa }(r)$ in Eq. (22b), we get two second-order non-linear
differential equations for the upper and lower radial spinor components,
respectively
\end{subequations}
\begin{equation*}
\left\{ \frac{d^{2}}{dr^{2}}-\frac{\kappa \left( \kappa +1\right) }{r^{2}}+%
\frac{2\kappa }{r}U(r)-\frac{dU(r)}{dr}-U^{2}(r)+\frac{\frac{d\Delta (r)}{dr}%
}{M+E_{n\kappa }-\Delta (r)}\left( \frac{d}{dr}+\frac{\kappa }{r}%
-U(r)\right) \right\} F_{n\kappa }(r)
\end{equation*}%
\begin{equation}
=\left( M+E_{n\kappa }-\Delta (r)\right) \left( M-E_{n\kappa }+\Sigma
(r)\right) F_{n\kappa }(r),
\end{equation}%
\begin{equation*}
\left\{ \frac{d^{2}}{dr^{2}}-\frac{\kappa \left( \kappa -1\right) }{r^{2}}+%
\frac{2\kappa }{r}U(r)+\frac{dU(r)}{dr}-U^{2}(r)+\frac{\frac{d\Sigma (r)}{dr}%
}{M-E_{n\kappa }+\Sigma (r)}\left( \frac{d}{dr}-\frac{\kappa }{r}%
+U(r)\right) \right\} G_{n\kappa }(r)
\end{equation*}%
\begin{equation}
=\left( M+E_{n\kappa }-\Delta (r)\right) \left( M-E_{n\kappa }+\Sigma
(r)\right) G_{n\kappa }(r),
\end{equation}%
where $\kappa \left( \kappa +1\right) =l(l+1)$ and $\kappa \left( \kappa
-1\right) =\widetilde{l}(\widetilde{l}+1).$ The radial wave functions are
required to satisfy the necessary boundary conditions, that is, $F_{n\kappa
}(0)=G_{n\kappa }(0)=0$ and $F_{n\kappa }(r)=G_{n\kappa }(r)\rightarrow 0$
at infinity$.$

At this stage, we take the $\Sigma (r)$ or $\Delta (r)$ the form of Hulth%
\'{e}n potential (1) and the tensor potential in the form of the
Coulomb-like interaction. Equations (23) and (24) can be solved exactly for $%
\kappa =0,-1$ and $\kappa =0,1,$ respectively, because of the spin-orbit
centrifugal term.

\subsection{Spin symmetric bound state solution}

In this part we are taking the $\Sigma (r)$ as the Hulth\'{e}n potential and
$\Delta (r)=C_{s}=$ constant $\left( \frac{d\Delta (r)}{dr}=0\right) ,$ i.e.,%
\begin{equation}
\Sigma (r)=V_{H}(r)=-\frac{\Sigma _{0}}{e^{r/r_{0}}-1},\text{ }\Sigma
_{0}=V_{0},
\end{equation}%
In fact, when the limit of $r_{0}$ becomes infinity, then $\underset{%
r_{0}\rightarrow \infty }{\lim }\Sigma (r)=\infty ,$ it tells us that the
Dirac particle could not be trapped by the Hulth\'{e}n potential, which does
not have any bound state under the condition of spin symmetry. The
Coulomb-like potential [58] for the tensor due to a charge $Z_{a}e$
interacting with a charge $Z_{A}e,$ distributed uniformly over a sphere of
radius $R_{c},$ is added,%
\begin{equation}
U_{Coul}(r)=-\frac{H}{r},\text{ }H=\frac{Z_{a}Z_{A}e^{2}}{4\pi \varepsilon
_{0}},\text{ }r\geq R_{c},
\end{equation}%
where $R_{c}=7.78$ fm is the Coulomb radius, and $Z_{a}$ and $Z_{A}$ denote
the charges of the projectile $a$ and the target nuclei $A,$ respectively.
Under this symmetry, substituting Eqs. (25) and (26) into Eq. (23), the
equation obtained for the upper radial spinor $F_{n\kappa }(r)$ becomes
\begin{equation}
\left[ \frac{d^{2}}{dr^{2}}-\frac{\gamma }{r^{2}}+\frac{\beta }{\left(
e^{r/r_{0}}-1\right) r_{0}^{2}}+\frac{\mathcal{E}_{nk}}{r_{0}^{2}}\right]
F_{n\kappa }(r)=0,
\end{equation}%
where
\begin{subequations}
\begin{equation}
\gamma =\eta _{k}\left( \eta _{k}-1\right) ,\text{ }\eta _{k}=\kappa +H+1,
\end{equation}%
\begin{equation}
\beta =r_{0}^{2}\left( E_{n\kappa }+M-C_{s}\right) V_{0}>0,
\end{equation}%
\begin{equation}
\mathcal{E}_{nk}=r_{0}^{2}\left( E_{n\kappa }-M\right) \left( E_{n\kappa
}+M-C_{s}\right) \leq \gamma d_{0},\text{ }M\geq E_{n\kappa },\text{ }
\end{equation}%
with $\kappa =l$ and $\kappa =-\left( l+1\right) $ for $\kappa >0$ and $%
\kappa <0,$ respectively$.$ Also, the quantum condition in (27) is obtained
from the finiteness of the solution at the origin point
\end{subequations}
\begin{equation}
F_{n\kappa }(0)=0,
\end{equation}%
and at infinity
\begin{equation}
\underset{r\rightarrow \infty }{\lim }F_{n\kappa }(r)=0,
\end{equation}%
The spin symmetric energy eigenvalues depend on $n$ and $\kappa ,$ \textit{%
i.e.}, $E_{n\kappa }=E(n,\kappa \left( \kappa +1\right) ).$ In Eq. (25), the
choice of $\Sigma (r)=2V(r)\rightarrow V(r)$ as mentioned in Ref. [12]
allows us to reduce the resulting relativistic solutions into their
non-relativistic limits under appropriate transformations.

Equation (27) can not be solved exactly only for the case of $\kappa =1$ due
to the spin-orbit centrifugal term $\gamma r^{-2}$ which is expanded in
terms of singular functions of e$^{-r/r_{0}}$ compatible with the
solvability of the problem for $r\ll r_{0}.$ Now, since the orbital term $%
r^{-2}$ is too singular, the validity of such approximation is limited only
to very few of the lowest energy states. Therefore, to go to higher energy
states one may attempt to solve the relativistic version of the problem in
[44] like the Dirac equation (27). Therefore, we apply the approximation
scheme derived in [44] for the centrifugal term which is valid for small
screening parameter $\delta $ values (i.e., for large $r_{0}$ values or $%
r\ll r_{0}).$ It can be casted in the form [44]:
\begin{equation}
\frac{1}{r^{2}}\approx \frac{1}{r_{0}^{2}}\left[ d_{0}+\frac{1}{e^{r/r_{0}}-1%
}+\frac{1}{\left( e^{r/r_{0}}-1\right) ^{2}}\right] ,
\end{equation}%
where the dimensionless constant $d_{0}=1/12$ is exact as reported
in many recent works (cf. e.g., [35,59-62])$.$ Obviously, the above
approximation to the centrifugal (pseudo-centrifugal) term turns to
$r^{-2}$ when the parameter $r_{0}$ goes to infinity (small
screening
parameter $\delta $) as%
\begin{equation*}
\underset{r_{0}\rightarrow \infty }{\lim }\left[ \frac{1}{r_{0}^{2}}\left(
d_{0}+\frac{1}{e^{r/r_{0}}-1}+\frac{1}{\left( e^{r/r_{0}}-1\right) ^{2}}%
\right) \right] =\frac{1}{r^{2}},
\end{equation*}%
which shows that the usual approximation is the limit of our approximation
(cf. e.g., [44] and the references therein). In terms of the new variable $%
z(r)=e^{-r/r_{0}},$ which maps the interval $r\in (0,\infty )$ into $z\in
(0,1),$ and using the approximation in Eq. (31)$,$ then Eq. (27) transforms
into rational functions as%
\begin{equation}
\left\{ \frac{d^{2}}{dz^{2}}+\frac{(1-z)}{z(1-z)}\frac{d}{dz}-\frac{1}{%
z^{2}(1-z)^{2}}\left[ q_{2}z^{2}-q_{1}z+q_{0}\right] \right\} F_{n,\kappa
}(z)=0,
\end{equation}%
where
\begin{subequations}
\begin{equation}
q_{2}=\beta +\gamma d_{0}-\mathcal{E}_{n\kappa },
\end{equation}%
\begin{equation}
q_{1}=\beta +2\gamma d_{0}-\gamma -2\mathcal{E}_{n\kappa },\text{ }
\end{equation}%
\begin{equation}
q_{0}=\gamma d_{0}-\mathcal{E}_{nk}=\lambda _{n\kappa }^{2},
\end{equation}%
and $\lambda _{n\kappa }$ must be a positive real parameter for the presence
of bound states. We are looking for solutions in the form
\end{subequations}
\begin{equation}
F_{n,\kappa }(z)=z^{\xi }(1-z)^{\varsigma }f(z),\text{ }\xi >0,\text{ }%
\varsigma \geq 1,
\end{equation}%
where $\xi $ and $\varsigma $ are real positive parameters and the boundary
conditions in Eqs. (29) and (30) are also satisfied. Then, by substituting
(34) into (32) the following hypergeometric differential equation for $f(z)$
is obtained%
\begin{equation}
z(1-z)\frac{d^{2}}{dz^{2}}f(z)+\left[ 2\xi +1-\left( 2\xi +2\varsigma
+1\right) z\right] \frac{d}{dz}f(z)-\left[ \left( \xi +\varsigma \right)
^{2}-q_{2}\right] f(z)=0,
\end{equation}%
Thus, the wave functions satisfying Eqs. (27), (30) and (34) are given by%
\begin{equation}
F_{n,\kappa }(z)=\mathcal{N}_{n\kappa }(1-z)^{\eta _{k}}z^{\lambda _{n\kappa
}}%
\begin{array}{c}
_{2}F_{1}%
\end{array}%
\left( -n,n+2\lambda _{n\kappa }+2\eta _{k}+1,2\lambda _{n\kappa
}+1;z\right) ,
\end{equation}%
where%
\begin{equation}
\eta _{k}=\frac{1}{2}+\sqrt{\frac{1}{4}+q_{0}+q_{2}-q_{1}}=\kappa
+H+1\rightarrow \eta _{k}^{\pm }=\left\{
\begin{array}{cc}
l+H+1, & \kappa >0, \\
-l+H, & \kappa <0.%
\end{array}%
\right.
\end{equation}%
For the bound-state problem, the solutions (36) fulfill the boundary
condition (29) when%
\begin{equation}
\lambda _{n\kappa }+\eta _{k}-\sqrt{q_{2}}=-n,\text{ }n=0,1,2,\cdots ,
\end{equation}%
where $n$ is the number of the nodes of the radial wave functions. This also
results in the following energy spectrum formula%
\begin{equation}
\lambda _{n\kappa }=\frac{\beta -\left( n+\kappa +H+1\right) ^{2}}{2\left(
n+\kappa +H+1\right) }.
\end{equation}%
The last formula gives an equation for the energy levels. When we insert $%
\lambda _{n\kappa }$ and $\beta $ into the above equation, we find the
following transcendental energy spectrum formula%
\begin{equation*}
\left( E_{n\kappa }-M\right) \left( E_{n\kappa }+M-C_{s}\right) =\frac{%
d_{0}\left( \kappa +H\right) \left( \kappa +H+1\right) }{r_{0}^{2}}
\end{equation*}%
\begin{equation*}
-\frac{1}{4}\left[ \frac{r_{0}\left( E_{n\kappa }+M-C_{s}\right) V_{0}}{%
\left( n+\kappa +H+1\right) }-\frac{\left( n+\kappa +H+1\right) }{r_{0}}%
\right] ^{2},
\end{equation*}%
\begin{equation}
n=0,1,2,\cdots ,n_{\max }=r_{0}\sqrt{V_{0}\left( E_{n\kappa }+M-C_{s}\right)
}-\kappa -H-1,
\end{equation}%
where $n_{\max }$ being the largest integer which is less than $r_{0}\sqrt{%
V_{0}\left( E_{n\kappa }+M-C_{s}\right) }-\kappa -H-1,$ $E_{n\kappa
}>C_{s}-M $. In the limit of zero tensor couplings ($H=0$), the spin aligned
states ($\kappa <0$) demand that $n+1\neq -\kappa .$ However, such a
condition is no longer exist for the spin unaligned states ($\kappa >0$).

On the other hand, we may also apply the NU method, this can be done when
Eq. (32) is compared with Eq. (14), the following polynomials are found%
\begin{equation}
\widetilde{\tau }(z)=1-z,\text{\ }\sigma (z)=z(1-z),\text{\ }\widetilde{%
\sigma }(z)=q_{2}z^{2}-q_{1}z+q_{0}.
\end{equation}%
Furthermore, we can follow the parametric generalization model of the NU
method given in Appendix A to obtain the specific values for the constants $%
c_{i}$ $(i=1,2,\cdots ,13),$ the results are listed in Table 1 for the
potential model under consideration. When the relations (A1-A4) of Appendix
A together with the values of constants given in Table 1 are applied, the
key polynomials take the following particular forms [63]:
\begin{equation}
\pi (z)=-\frac{z}{2}-\left[ \left( \eta _{k}+\lambda _{n\kappa }-\frac{1}{2}%
\right) z-\lambda _{n\kappa }\right] ,
\end{equation}%
and%
\begin{equation}
k=\beta -\gamma -2\lambda _{n\kappa }\left( \eta _{k}-\frac{1}{2}\right) ,
\end{equation}%
for discrete bound state solutions. According to the NU method, we further
obtain%
\begin{equation*}
\tau (z)=1-2\left[ \left( \eta _{k}+\lambda _{n\kappa }+\frac{1}{2}\right)
z-\lambda _{n\kappa }\right] ,
\end{equation*}%
\begin{equation}
\text{ }\tau ^{\prime }(z)=-2\left( \eta _{k}+\lambda _{n\kappa }+\frac{1}{2}%
\right) <0,
\end{equation}%
with prime denotes the derivative with respect to $z.$ Referring to Table 1
and applying the relation A5 of Appendix A, we obtain energy equation
formula which is identical to Eq. (40).

In the limit of zero tensor coupling and under the usual approximation, $H=$
$d_{0}=C_{s}=0,$ $\kappa =l$ and $V_{0}=Ze^{2}\delta ,$ Eq. (40) becomes%
\begin{equation*}
\sqrt{M^{2}-E_{nl}^{2}}=\frac{Ze^{2}\left( E_{nl}+M\right) }{2\left(
n+l+1\right) }-\frac{\left( n+l+1\right) }{2}\delta ,\text{ }\left\vert
E_{nl}\right\vert <M,
\end{equation*}%
which is completely identical to Eq. (68) of Ref. [64] for the solution of
the Klein-Gordon equation with equally mixed scalar and vector Hulth\'{e}n
potentials, i.e., $V_{0}=S_{0}$ ($\Sigma _{0}=0$). Furthermore, the last
equation is also identical to Eq. (25) upon inserting $q=1,$ $V_{0}=S_{0}$
and $\delta =\delta _{\pm }=l+1,-l$ in Eqs. (8) and (9) of Ref. [65].
Therefore, real solutions are possible only for $\left\vert
E_{nl}\right\vert \leq M$ (i.e., bound states).

We now look at some special cases and relationships between our results and
some other existing results in literature. For exact spin symmetry, $%
V_{0}=S_{0}$ case or $C_{s}=0$ and applying the following appropriate
transformations $E_{nl}+M\simeq 2\mu ,$ $E_{nl}-M\simeq E_{nl}$ and $\kappa
=l$, we obtain the following non-relativistic energy spectrum for the Hulth%
\'{e}n potential including the Coulomb-like interaction,
\begin{equation}
E_{nl}=\frac{\delta ^{2}}{2\mu }\left\{ \left( H+l\right) \left(
H+l+1\right) d_{0}-\left[ \frac{\mu Ze^{2}}{\delta \left( H+n+l+1\right) }-%
\frac{\left( H+n+l+1\right) }{2}\right] ^{2}\right\} ,
\end{equation}%
leading to the following energy spectrum formula in the limit of zero tensor
coupling [44]%
\begin{equation*}
E_{nl}=\frac{\delta ^{2}}{2M}\left\{ l\left( l+1\right) d_{0}-\left[ \frac{%
\mu Ze^{2}}{\delta \left( n+l+1\right) }-\frac{\left( n+l+1\right) }{2}%
\right] ^{2}\right\} ,
\end{equation*}%
\begin{equation}
n=0,1,2,\cdots ,n_{\max }=r_{0}\sqrt{2\mu V_{0}}-l-1,
\end{equation}%
which is identical to Eq. (34) in Ref. [44] for the \ solution of the Schr%
\"{o}dinger equation with the Hulth\'{e}n potential. The largest integer $%
n_{\max }\leq r_{0}\sqrt{2\mu V_{0}}-l-1$ which is identical to Eqs. (12)
and (13) given in Ref. [66] after choosing $A=2\mu r_{0}^{2}V_{0},$ $\alpha
=1$ and $\alpha =0$ for the Hulth\'{e}n potential case$,$ respectively. In
fact, when the limit of $\delta $ becomes zero, then%
\begin{equation*}
\underset{r_{0}\rightarrow \infty }{\lim }E_{nl}=-\frac{1}{2M}\frac{\left(
\mu Ze^{2}\right) ^{2}}{\left( n+l+1\right) ^{2}},
\end{equation*}%
which is a spectrum resulting from a Coulombic field.

We apply now the relations (A6-A10) of Appendix A to calculate the
corresponding wave functions as [63]%
\begin{equation}
\rho (z)=z^{2\lambda _{n\kappa }}(1-z)^{2\eta _{\kappa }},
\end{equation}%
\begin{equation}
\phi (z)=z^{\lambda _{n\kappa }}(1-z)^{\eta _{\kappa }},
\end{equation}%
where $\eta _{\kappa }\geq 1$ and $\lambda _{n\kappa }$ must be a positive
real parameter for the presence of bound states. Hence, we find%
\begin{equation}
y_{n}(z)=P_{n}^{\left( 2\lambda _{n\kappa },2\eta _{\kappa }\right) }(1-2z),%
\text{ }z\in \lbrack 0,1],
\end{equation}%
with $P_{n}^{\left( \nu ,\mu \right) }(x)$ is the Jacobi polynomial, where $%
\nu >-1,$ $\mu >-1$ and defined for the argument $x\in \lbrack -1,1].$ Using
the relation $F_{n\kappa }(z)=\phi (z)y_{n}(z),$ we get the radial
upper-spinor wave functions as%
\begin{equation}
F_{n\kappa }(r)=\mathcal{N}_{n\kappa }\frac{n!\Gamma (2\lambda _{n\kappa }+1)%
}{\Gamma (n+2\lambda _{n\kappa }+1)}e^{-\left( \lambda _{n\kappa
}/r_{0}\right) r}\left( 1-e^{-r/r_{0}}\right) ^{\eta _{\kappa
}}P_{n}^{\left( 2\lambda _{n\kappa },2\eta _{\kappa }\right)
}(1-2e^{-r/r_{0}}).
\end{equation}%
Furthermore, we give the relation linking the hypergeometric function and
the Jacobi polynomials (see formula 8.962.1) in [67]%
\begin{equation}
\begin{array}{c}
_{2}F_{1}%
\end{array}%
\left( -n,n+\nu +\mu +1,\nu +1;\frac{1-x}{2}\right) =\frac{n!\Gamma (\nu +1)%
}{\Gamma (n+\nu +1)}P_{n}^{\left( \nu ,\mu \right) }(x),
\end{equation}%
to rewrite the radial wave functions as%
\begin{equation}
F_{n\kappa }(r)=\mathcal{N}_{n\kappa }e^{-\left( \lambda _{n\kappa
}/r_{0}\right) r}\left( 1-e^{-r/r_{0}}\right) ^{\eta _{\kappa }}%
\begin{array}{c}
_{2}F_{1}%
\end{array}%
\left( -n,n+2\lambda _{n\kappa }+2\eta _{\kappa }+1,2\lambda _{n\kappa
}+1;e^{-r/r_{0}}\right) .
\end{equation}%
where the normalization constant is calculated in Appendix B in closed form.

Before presenting the corresponding lower-component $G_{n\kappa }(r),$ let
us recall a recurrence relation of hypergeometric function,%
\begin{equation}
\frac{d}{dz}\left[
\begin{array}{c}
_{2}F_{1}%
\end{array}%
\left( a;b;c;z\right) \right] =\left( \frac{ab}{c}\right)
\begin{array}{c}
_{2}F_{1}%
\end{array}%
\left( a+1;b+1;c+1;z\right) ,
\end{equation}%
where%
\begin{equation}
\begin{array}{c}
_{2}F_{1}%
\end{array}%
\left( a,b;c;z\right) =\frac{\Gamma (c)}{\Gamma (a)\Gamma (b)}%
\sum\limits_{p=0}^{\infty }\frac{\Gamma (a+p)\Gamma (b+p)}{\Gamma (c+p)}%
\frac{z^{p}}{p!}.
\end{equation}%
which is used to solve Eq. (22a) and present the corresponding lower
component $G_{n\kappa }(r)$ as follows%
\begin{equation*}
G_{n\kappa }(r)=\frac{1}{M+E_{n\kappa }-C_{s}}\left[ \frac{d}{dr}+\frac{%
\kappa }{r}-U(r)\right] F_{n\kappa }(r)
\end{equation*}%
\begin{equation*}
=\frac{\mathcal{N}_{n\kappa }e^{-\left( \lambda _{n\kappa }/r_{0}\right)
r}\left( 1-e^{-r/r_{0}}\right) ^{\eta _{\kappa }}}{r_{0}\left( M+E_{n\kappa
}-C_{s}\right) }\left\{ \left[ -\lambda _{n\kappa }+\eta _{\kappa }\frac{%
e^{-r/r_{0}}}{1-e^{-r/r_{0}}}+\frac{\left( \kappa +H\right) r_{0}}{r}\right]
\right.
\end{equation*}%
\begin{equation*}
\times
\begin{array}{c}
_{2}F_{1}%
\end{array}%
\left( -n,n+2\lambda _{n\kappa }+2\eta _{\kappa }+1,2\lambda _{n\kappa
}+1;e^{-r/r_{0}}\right)
\end{equation*}%
\begin{equation}
-\left.
\begin{array}{c}
_{2}F_{1}%
\end{array}%
\left( -n+1,n+2\left( \lambda _{n\kappa }+\eta _{\kappa }+1\right) ,2\left(
\lambda _{n\kappa }+1\right) ;e^{-r/r_{0}}\right) \right\} .
\end{equation}%
Here, we remark that the hypergeometric series $%
\begin{array}{c}
_{2}F_{1}%
\end{array}%
\left( -n,n+2\lambda _{n\kappa }+2\eta _{\kappa }+1,2\lambda _{n\kappa
}+1;e^{-r/r_{0}}\right) $ terminates for $n=0$ and thus converges for all
values of real parameters $\eta _{\kappa }$ and $\lambda _{n\kappa }.$ It is
worthy to note that in the limit of $r_{0}$ goes to infinity then the Dirac
spinor components become as%
\begin{equation*}
\underset{r_{0}\rightarrow \infty }{\lim }F_{n\kappa }(r)=\underset{%
r_{0}\rightarrow \infty }{\lim }G_{n\kappa }(r)=0,
\end{equation*}%
which show that these components become unbound.

For $C_{s}>M+E_{n\kappa }$ and $E_{n\kappa }<M$ or $C_{s}<M+E_{n\kappa }$
and $E_{n\kappa }>M,$ we note that parameters given in Eq. (33c) turn to be
imaginary, i.e., $\lambda _{n\kappa }^{2}<0$ in the $s$-state ($\kappa =-1$)
case$.$ As a result, the condition of existing bound states are $\lambda
_{n\kappa }>0$ and $\eta _{\kappa }>0,$ that is to say, in the case of $%
C_{s}>M+E_{n\kappa }$ and $E_{n\kappa }<M,$ bound-states do not exist for
some quantum number $\kappa $ such as the $s$-state ($\kappa =-1$)$.$ Of
course, if these conditions are satisfied for existing bound-states, the
energy equation and wave functions are the same as these given in Eqs. (45),
(52) and (55).

In the spin-symmetric case, the numerical bound state energy spectrum is
obtained from Eq. (40) with the choice of the following values for the
parameters: $M=10$ $fm^{-1},$ $r_{0}=10$ $fm,$ $V_{0}=10$ $fm^{-1}$ and $%
C_{s}=10.1$ $fm^{-1}.$ Thus the approximated values for the energy states
using the usual approximation scheme (31) (i.e., with $d_{0}=0)$ are
presented in Table 2 for states $n=0,1,2$ and $3$ and spin-orbit quantum
number $\kappa =\pm 1,\pm 2,\pm 3$ and $\pm 4$ values. Apparantly, the spin
unaligned states $(\kappa >0),$ the orbital states $(l=\left\vert \kappa
+1/2\right\vert -1/2,$ $j=\left\vert \kappa \right\vert -1/2)$: $%
(1p_{1/2},0d_{3/2}),$ $(2p_{1/2},1d_{3/2},0f_{5/2}),$ $%
(3p_{1/2},2d_{3/2},1f_{5/2})$ and $(3d_{3/2},2f_{5/2})$ are found to be
degenerate in the presence of tensor $(H\neq 0)$ and in the limit of zero
tensor couplings ($H=0)$. In addition, the energies of the $np_{1/2},$ $%
nd_{3/2}$ and $nf_{5/2}$ states for $H\neq 0$ are higher than the
corresponding values for $H=0.$ Overmore, we reproduced the energy spectrum
of Table 1 using the new approximation scheme (31) proposed in Ref. [44]
through taking the dimensionless constant $d_{0}\approx 1/12$ [35,59-62].
The energy levels presented in Table 2 are slightly different under our new
approximation scheme. Further, the degeneracies in the above states have no
lnoger exist. This is true for both Coulomb tensor $(H\neq 0)$ and in the
limit of zero tensor ($H=0)$ interactions.

\subsection{Pseudospin symmetric bound state solutions}

The exact pseudospin symmetry occurs in the Dirac equation when $S(r)\sim
-V(r)$. In this part we will take the Hulth\'{e}n potential for $\Delta (r),$
(\textit{i.e}., $\frac{d\Sigma (r)}{dr}=0,$ or $\Sigma (r)=C_{ps}=$ constant)%
$,$%
\begin{equation}
\Delta (r)=-\frac{\Delta _{0}}{e^{r/r_{0}}-1},\text{ }\Delta _{0}=V_{0},
\end{equation}%
where $\Delta _{0}$ is a constant and the tensor potential as in Eq. (26).
Thus, Eq. (24) can be reduced into the Schr\"{o}dinger-type:
\begin{equation}
\left[ \frac{d^{2}}{dr^{2}}-\frac{\widetilde{\gamma }}{r^{2}}+\frac{%
\widetilde{\beta }}{\left( e^{r/r_{0}}-1\right) r_{0}^{2}}+\frac{\widetilde{%
\mathcal{E}}_{nk}}{r_{0}^{2}}\right] F_{n\kappa }(r)=0,
\end{equation}%
where
\begin{subequations}
\begin{equation}
\widetilde{\gamma }=\Lambda _{k}\left( \Lambda _{k}-1\right) ,\text{ }%
\Lambda _{k}=\kappa +H,
\end{equation}%
\begin{equation}
\widetilde{\beta }=r_{0}^{2}\left( E_{n\kappa }-M-C_{ps}\right) V_{0},
\end{equation}%
\begin{equation}
\widetilde{\mathcal{E}}_{nk}=r_{0}^{2}\left( E_{n\kappa }+M\right) \left(
E_{n\kappa }-M-C_{ps}\right) \leq \gamma d_{0},\text{ }E_{n\kappa }\leq
M+C_{ps},\text{ }
\end{equation}%
where $\kappa =-\widetilde{l}$ and $\kappa =\widetilde{l}+1$ for $\kappa <0$
and $\kappa >0,$ respectively. \ In the pseudospin symmetry, the eigenstates
with with $\widetilde{j}=\widetilde{l}\pm \frac{1}{2}$ are degenerate for $%
\widetilde{l}\neq 0.$ The energy eigenvalues $E_{n\kappa }$ depend only on $%
n $ and $\kappa ,$ \textit{i.e.}, $E_{n\kappa }=E(n,\kappa (\kappa -1)).$ We
now follow the previous procedures to obtain a differential equation for $%
G_{n,\kappa }(z),$
\end{subequations}
\begin{equation}
\left\{ \frac{d^{2}}{dz^{2}}+\frac{(1-z)}{z(1-z)}\frac{d}{dz}-\frac{1}{%
z^{2}(1-z)^{2}}\left[ p_{2}z^{2}-p_{1}z+p_{0}\right] \right\} G_{n,\kappa
}(z)=0,
\end{equation}%
where
\begin{subequations}
\begin{equation}
p_{2}=\widetilde{\beta }+\widetilde{\gamma }d_{0}-\widetilde{\mathcal{E}}%
_{n\kappa },
\end{equation}%
\begin{equation}
p_{1}=\widetilde{\beta }+2\widetilde{\gamma }d_{0}-\widetilde{\gamma }-2%
\widetilde{\mathcal{E}}_{n\kappa },\text{ }
\end{equation}%
\begin{equation}
p_{0}=\widetilde{\gamma }d_{0}-\widetilde{\mathcal{E}}_{nk}=\widetilde{%
\lambda }_{n\kappa }^{2},
\end{equation}%
and $\widetilde{\lambda }_{n\kappa }$ must be a positive real parameter for
real solution. Thus, the energy spectrum is then given by
\end{subequations}
\begin{equation}
\lambda _{n\kappa }=\frac{\widetilde{\beta }-\left( n+\kappa +H\right) ^{2}}{%
2\left( n+\kappa +H\right) }.
\end{equation}%
The last formula gives an equation for the energy spectrum. When we insert $%
\widetilde{\lambda }_{n\kappa }$ and $\widetilde{\beta }$ into the above
equation, then we find the following transcendental energy equation%
\begin{equation*}
\left( E_{n\kappa }+M\right) \left( E_{n\kappa }-M-C_{ps}\right) =\frac{%
d_{0}\left( \kappa +H\right) \left( \kappa +H-1\right) }{r_{0}^{2}}
\end{equation*}%
\begin{equation*}
-\frac{1}{4}\left[ \frac{r_{0}\left( E_{n\kappa }-M-C_{ps}\right) V_{0}}{%
\left( n+\kappa +H\right) }-\frac{\left( n+\kappa +H\right) }{r_{0}}\right]
^{2},
\end{equation*}%
\begin{equation}
n=0,1,2,\cdots ,n_{\max }=r_{0}\sqrt{V_{0}\left( E_{n\kappa }-M-C_{s}\right)
}-\kappa -H,
\end{equation}%
where $n_{\max }$ being the largest integer which is less than $r_{0}\sqrt{%
V_{0}\left( E_{n\kappa }-M-C_{s}\right) }-\kappa -H,$ where $E_{n\kappa
}>M+C_{s}$. Obviously, the above equation can be reduced to Klein-Gordon
solution when $C_{ps}=H=d_{0}=0$ and after inserting $q=1,$ $V_{0}=S_{0}$
and $\delta =\delta _{\pm }=\widetilde{l}+1,-\widetilde{l}$ in Eqs. (8), (9)
and (25) of Ref. [65]. Furthermore, the wave functions can be written as%
\begin{equation*}
G_{n,\kappa }(z)=\mathcal{N}_{n\kappa }(1-z)^{\Lambda _{k}}z^{\widetilde{%
\lambda }_{n\kappa }}%
\begin{array}{c}
_{2}F_{1}%
\end{array}%
\left( -n,n+2\widetilde{\lambda }_{n\kappa }+2\Lambda _{k}+1,2\widetilde{%
\lambda }_{n\kappa }+1;z\right) ,
\end{equation*}%
\begin{equation}
=\mathcal{N}_{n\kappa }\frac{n!\Gamma (2\widetilde{\lambda }_{n\kappa }+1)}{%
\Gamma (n+2\widetilde{\lambda }_{n\kappa }+1)}e^{-\left( \widetilde{\lambda }%
_{n\kappa }/r_{0}\right) r}\left( 1-e^{-r/r_{0}}\right) ^{\Lambda _{\kappa
}}P_{n}^{\left( 2\widetilde{\lambda }_{n\kappa },2\Lambda _{\kappa }\right)
}(1-2e^{-r/r_{0}}).
\end{equation}%
where%
\begin{equation}
\Lambda _{\kappa }=\kappa +H,\text{ }\Lambda _{k}^{\pm }=\left\{
\begin{array}{cc}
\widetilde{l}+H+1, & \kappa >0, \\
-\widetilde{l}+H, & \kappa <0.%
\end{array}%
\right.
\end{equation}%
To avoid repetition, the solution of Eq. (24) can be found easily by
applying appropriate paremetric transformations. A first inspection for the
relationship between the present set of parameters $(\widetilde{\mathcal{E}}%
_{nk},\widetilde{\beta },\Lambda _{k})$ and the previous set $(\mathcal{E}%
_{nk},\beta ,\eta _{\kappa })$ tells us that the energy solution, for
pseudospin symmetry, can be obtained directly from those of the previous
energy solution, in spin symmetric case, by applying the following
appropriate parameter map [42,43]:
\begin{equation}
F_{n,\kappa }(r)\leftrightarrow G_{n,\kappa }(r),\text{ }E_{n,\kappa
}\rightarrow -E_{n,\kappa },\text{ }C_{s}\rightarrow -C_{ps},\text{ }%
V(r)\rightarrow -V(r)\text{ }(\text{or }V_{0}\rightarrow -V_{0}),
\end{equation}%
or simply we apply the following transformations:%
\begin{equation}
F_{n,\kappa }(r)\leftrightarrow G_{n,\kappa }(r),\text{ }M\rightarrow -M,%
\text{ }C_{s}\rightarrow C_{ps},\text{ }\eta _{\kappa }\rightarrow \Lambda
_{\kappa }\text{ }(\text{or }\kappa +H+1\rightarrow \kappa +H,\text{ }\kappa
+H\rightarrow \kappa +H-1),
\end{equation}%
on Eqs. (40) and (50) to obtain Eqs. (62) and (63) for energy spectrum
formula and wave functions, respectively.

\section{Concluding Remarks}

In this work, we have obtained analytically the approximate energy equation
and the corresponding wavefunctions of the Dirac equation for the Hulth\'{e}%
n potential coupled with a Coulombic-like tensor under the conditions of the
spin and pseudospin symmetry using the parametric generalization of the NU
method. The introduced Coulombic-like tensor interaction generates
additional centrifugal-like term $L(L+1)r^{-2}.$ Therefore, in order to
solve the resulting Schr\"{o}dinger-like equation analytically, a new
approximation scheme was used to approximate the new spin-coupling
centrifugal term $\gamma r^{-2}$ arising from the Coulomb-like tensor which
yields a closed form solution of the problem under consideration. The
resulting solutions of the wavefunctions are written in terms of the
orthogonal Jacobi polynomials or hypergeometric functions. Obviously, for
exact spin when $S(r)=V(r)$ (\textit{i.e}., $C_{s}=0$), the relativistic
solution can be easily reduced to it's non-relativistic limit by the choice
of appropriate mapping transformations. Also, in the limit of zero tensor
couplings, the present results reduce to the well-known solutions of the
Dirac equation for the usual Hulth\'{e}n potential with arbitrary spin-orbit
coupling quantum number $\kappa $.

\acknowledgments The authors highly appreciate the very constructive
comments and suggestions provided from the kind referee and editor. The
support provided by the Scientific and Technological Research Council of
Turkey (T\"{U}B\.{I}TAK) is highly appreciated.

\newpage \appendix

\section{Parametric Generalization of the NU Method}

We complement the theoretical parameterized formulation of the NU method for
any arbitrary exponential potential by giving the essential polynomials,
energy equation and wavefunctions together with their relevant constants as
follows [42,43,63].

(i) The key NU polynomials:
\begin{equation}
\pi (z)=c_{4}+c_{5}z-\left[ \left( \sqrt{c_{9}}+c_{3}\sqrt{c_{8}}\right) z-%
\sqrt{c_{8}}\right] ,
\end{equation}%
\begin{equation}
k=-\left( c_{7}+2c_{3}c_{8}\right) -2\sqrt{c_{8}c_{9}}.
\end{equation}%
\begin{equation}
\tau (z)=1-\left( c_{2}-2c_{5}\right) z-2\left[ \left( \sqrt{c_{9}}+c_{3}%
\sqrt{c_{8}}\right) z-\sqrt{c_{8}}\right] ,
\end{equation}%
\begin{equation}
\tau ^{\prime }(z)=-2c_{3}-2\left( \sqrt{c_{9}}+c_{3}\sqrt{c_{8}}\right) <0,
\end{equation}%
(ii) The general energy equation:%
\begin{equation}
\left( c_{2}-c_{3}\right) n+c_{3}n^{2}-\left( 2n+1\right) c_{5}+\left(
2n+1\right) \left( \sqrt{c_{9}}+c_{3}\sqrt{c_{8}}\right) +c_{7}+2c_{3}c_{8}+2%
\sqrt{c_{8}c_{9}}=0.
\end{equation}%
(iii) The general wavefunctions:%
\begin{equation}
\rho (z)=z^{c_{10}}(1-c_{3}z)^{c_{11}},
\end{equation}%
\begin{equation}
\phi (z)=z^{c_{12}}(1-c_{3}z)^{c_{13}},\text{ }c_{12}>0,\text{ }c_{13}>0,
\end{equation}%
\begin{equation}
y_{n}(z)=P_{n}^{\left( c_{10},c_{11}\right) }(1-2c_{3}z),\text{ }\mathbb{R}%
c_{10}>-1,\text{ }\mathbb{R}c_{11}>-1,
\end{equation}%
\begin{equation}
u(z)=\mathcal{N}_{n}z^{c_{12}}(1-c_{3}z)^{c_{13}}P_{n}^{\left(
c_{10},c_{11}\right) }(1-2c_{3}z),
\end{equation}%
where $P_{n}^{\left( \nu ,\mu \right) }(x),$ $\mathbb{R}$$\nu >-1,$
$\mathbb{R}\mu >-1$
and $x\in \lbrack -1,1]$ are the Jacobi polynomials and $%
\mathcal{N}_{n}$ is a normalization constants. Also, the above wave
functions can be expressed in terms of the hypergeometric function as%
\begin{equation}
u_{n\kappa }(z)=\mathcal{N}_{n\kappa }z^{c_{12}}(1-c_{3}z)^{c_{13}}%
\begin{array}{c}
_{2}F_{1}%
\end{array}%
\left( -n,1+c_{10}+c_{11}+n;c_{10}+1;c_{3}z\right) ,
\end{equation}%
where $c_{12}>0,$ $c_{13}>0$ and $z\in \left[ 0,1/c_{3}\right] .$

(iv) The relevant constants:%
\begin{equation*}
c_{4}=\frac{1}{2}\left( 1-c_{1}\right) ,\text{ }c_{5}=\frac{1}{2}\left(
c_{2}-2c_{3}\right) ,\text{ }c_{6}=c_{5}^{2}+B_{1},
\end{equation*}%
\begin{equation*}
\text{ }c_{7}=2c_{4}c_{5}-B_{2},\text{ }c_{8}=c_{4}^{2}+B_{3},\text{ }%
c_{9}=c_{3}\left( c_{7}+c_{3}c_{8}\right) +c_{6},
\end{equation*}%
\begin{equation*}
c_{10}=c_{1}+2c_{4}+2\sqrt{c_{8}}-1>-1,\text{ }c_{11}=1-c_{1}-2c_{4}+\frac{2%
}{c_{3}}\sqrt{c_{9}}>-1,
\end{equation*}%
\begin{equation}
c_{12}=c_{4}+\sqrt{c_{8}}>0,\text{ }c_{13}=-c_{4}+\frac{1}{c_{3}}\left(
\sqrt{c_{9}}-c_{5}\right) >0.
\end{equation}%
$\label{appendix}$

\section{Normalization of the radial wavefunctions}

The normalization constant $\mathcal{N}_{n\kappa }$ can be easily calculated
in closed form. To do this, we start by using the normalization condition $%
\int_{0}^{\infty }\left[ F_{n\kappa }(r)\right] ^{2}dr=1,$ and under the
coordinate change $x=1-2e^{-r/r_{0}},$ the normalization constant $\mathcal{N%
}_{n\kappa }$ in (50) is given by%
\begin{equation}
\mathcal{N}_{n\kappa }^{-2}=\frac{r_{0}}{2}\left[ \frac{n!\Gamma (2\lambda
_{n\kappa }+1)}{\Gamma (n+2\lambda _{n\kappa }+1)}\right] ^{2}\int_{-1}^{1}%
\left( \frac{1-x}{2}\right) ^{2\lambda _{n\kappa }-1}\left( \frac{1+x}{2}%
\right) ^{2\eta _{\kappa }}\left[ P_{n}^{(2\lambda _{n\kappa },2\eta
_{\kappa })}(x)\right] ^{2}dx,
\end{equation}%
where $\lambda _{n\kappa }>0$ and $\eta _{\kappa }\geq 1.$ The calculation
of this integral can be done by writting one of the $2\eta _{\kappa }$
factors $\left( 1+x\right) /2$ in the form%
\begin{equation}
\frac{1+x}{2}=1-\frac{1-x}{2},
\end{equation}%
and by making use of the following integral (see formula (7.391.5) in [67]):%
\begin{equation}
\int_{-1}^{1}\left( 1-x\right) ^{\nu -1}\left( 1+x\right) ^{\mu }\left[
P_{n}^{(\nu ,\mu )}(x)\right] ^{2}dx=\frac{2^{\nu +\mu }\Gamma (n+\nu
+1)\Gamma (n+\mu +1)}{n!\nu \Gamma (n+\nu +\mu +1)},
\end{equation}%
which is valid for $\mathbb{R}\nu >0$ and $\mathbb{R}\mu >-1.$ This
leads to

\begin{equation}
\mathcal{N}_{n\kappa }=\frac{1}{\Gamma (2\lambda _{n\kappa }+1)}\left[ \frac{%
2}{r_{0}}\frac{\lambda _{n\kappa }\Gamma (n+2\lambda _{n\kappa }+1)\Gamma
(n+2\lambda _{n\kappa }++2\eta _{\kappa }+1)}{n!\Gamma (n+2\eta _{\kappa }+1)%
}\right] ^{1/2}.
\end{equation}%
\newpage

\ {\normalsize 
}

\bigskip

\baselineskip= 2\baselineskip
\bigskip \newpage

\bigskip

\bigskip {\normalsize 
}

\baselineskip= 2\baselineskip

\bigskip
\begin{table}[tbp]
\caption{The NU parametric constants useful in calculating the energy
eigenvalues and eigenfunctions for the spin-symmetric Dirac equation.}%
\begin{tabular}{llll}
\tableline Constant & Analytic value & Constant & Analytic value \\
\tableline$c_{1}$ & 1 & $c_{2}$ & $1$ \\
$c_{3}$ & $1$ & c$_{4}$ & $0$ \\
$c_{5}$ & $-1/2$ & $c_{6}$ & $q_{2}+1/4$ \\
$c_{7}$ & $-q_{1}$ & $c_{8}$ & $\lambda _{n\kappa }^{2}$ \\
$c_{9}$ & $\left( \eta _{\kappa }-1/2\right) ^{2}$ & $c_{10}$ & $2\lambda
_{n\kappa }$ \\
$c_{11}$ & $2\eta _{\kappa }-1$ & $c_{12}$ & $\lambda _{n\kappa }$ \\
$c_{13}$ & $\eta _{\kappa }$ &  &  \\
\tableline &  &  &
\end{tabular}%
\end{table}
\

\bigskip

\begin{table}[tbp]
\caption{The usual approximation to the relativistic energy spectrum (in $%
fm^{-1})$ of the spin-symmetric Hulth\'{e}n potential including the Coulomb
coupling tensor of strength $H=0$ and $H=0.5$ for various values of $n$ and $%
\protect\kappa .$}%
\begin{tabular}{llllllllll}
\tableline$n$ & $l$ & $\kappa <0$ & $(l,j)$ & $E_{n,\kappa <0}(H=0)$%
\tablenotemark[1]\tablenotetext[1]{We have taken $d_{0}=0$ during these
calculations.} & $E_{n,\kappa <0}(H=0.5)$\tablenotemark[1] & $\kappa >0$ & $%
(l,j)$ & $E_{n,\kappa >0}(H=0)$\tablenotemark[1] & $E_{n,\kappa >0}(H=0.5)$%
\tablenotemark[1] \\
\tableline\tableline$0$ & $0$ & $-1$ & 0s$_{1/2}$ & $-$ & $0.101446652$ & $-$
& $-$ & $-$ & $-$ \\
$1$ & $0$ & $-1$ & 1s$_{1/2}$ & $0.1057848200$ & $0.100388800$ & $-$ & $-$ &
$-$ & $-$ \\
$2$ & $0$ & $-1$ & 2s$_{1/2}$ & $0.1231107030$ & $0.136077055$ & $-$ & $-$ &
$-$ & $-$ \\
$3$ & $0$ & $-1$ & $3s_{1/2}$ & $0.1518922615$ & $0.170536938$ & $-$ & $-$ &
$-$ & $-$ \\
$0$ & $1$ & $-2$ & $0p_{3/2}$ & $0.1057848200$ & $0.101446652$ & 1 & $%
0p_{1/2}$ & $0.123110703$ & $0.1360770550$ \\
$1$ & $1$ & $-2$ & $1p_{3/2}$ & $-$ & $0.101446652$ & 1 & $1p_{1/2}$ & $%
0.1518922615$ & $0.1705369375$ \\
$2$ & $1$ & $-2$ & $2p_{3/2}$ & $0.105784820$ & $0.100388800$ & 1 & $%
2p_{1/2} $ & $0.191988313$ & $0.2162202946$ \\
$3$ & $1$ & $-2$ & $3p_{3/2}$ & $0.123110703$ & $0.136077055$ & 1 & $%
3p_{1/2} $ & $0.243203547$ & $0.2729055754$ \\
$0$ & $2$ & $-3$ & $0d_{5/2}$ & $0.123110703$ & $0.100388800$ & 2 & $%
0d_{3/2} $ & $0.1518922615$ & $0.1705369375$ \\
$1$ & $2$ & $-3$ & $1d_{5/2}$ & $0.105784820$ & $0.101446652$ & 2 & $%
1d_{3/2} $ & $0.191988313$ & $0.2162202946$ \\
$2$ & $2$ & $-3$ & $2d_{5/2}$ & $-$ & $0.101446652$ & 2 & $2d_{3/2}$ & $%
0.243203547$ & $0.2729055754$ \\
$3$ & $2$ & $-3$ & $3d_{5/2}$ & $0.105784820$ & $0.100388800$ & 2 & $%
3d_{3/2} $ & $0.3052908168$ & $0.3403207458$ \\
$0$ & $3$ & $-4$ & $0f_{7/2}$ & $0.1518922615$ & $0.136077055$ & 3 & $%
0f_{5/2}$ & $0.191988313$ & $0.2162202946$ \\
$1$ & $3$ & $-4$ & $1f_{7/2}$ & $0.123110703$ & $0.100388800$ & 3 & $%
1f_{5/2} $ & $0.243203547$ & $0.2729055754$ \\
$2$ & $3$ & $-4$ & $2f_{7/2}$ & $0.105784820$ & $0.101446652$ & 3 & $%
2f_{5/2} $ & $0.3052908168$ & $0.3403207458$ \\
$3$ & $3$ & $-4$ & $3f_{7/2}$ & $-$ & $0.101446652$ & 3 & $3f_{5/2}$ & $%
0.3779539814$ & $0.4181464025$ \\
\tableline\tableline &  &  &  &  &  &  &  &  &
\end{tabular}%
\end{table}
\
\begin{table}[tbp]
\caption{The new approximation to the relativistic energy spectrum (in $%
fm^{-1})$ of the spin-symmetric Hulth\'{e}n potential including the Coulomb
coupling tensor of strength $H=0$ and $H=0.5$ for various values of $n$ and $%
\protect\kappa .$}%
\begin{tabular}{llllllllll}
\tableline$n$ & $l$ & $\kappa <0$ & $(l,j)$ & $E_{n,\kappa <0}(H=0)$%
\tablenotemark[1]\tablenotetext[1]{We have taken $d_{0}=0.0823058167837972$
during these calculations.} & $E_{n,\kappa <0}(H=0.5)$\tablenotemark[1] & $%
\kappa >0$ & $(l,j)$ & $E_{n,\kappa >0}(H=0)$\tablenotemark[1] & $%
E_{n,\kappa >0}(H=0.5)$\tablenotemark[1] \\
\tableline\tableline$0$ & $0$ & $-1$ & 0s$_{1/2}$ & $-$ & $0.1014318359$ & $%
- $ & $-$ & $-$ & $-$ \\
$1$ & $0$ & $-1$ & 1s$_{1/2}$ & $0.105784820$ & $0.1004034771$ & $-$ & $-$ &
$-$ & $-$ \\
$2$ & $0$ & $-1$ & 2s$_{1/2}$ & $0.123110703$ & $0.1360623873$ & $-$ & $-$ &
$-$ & $-$ \\
$3$ & $0$ & $-1$ & $3s_{1/2}$ & $0.1518922615$ & $0.1705222701$ & $-$ & $-$
& $-$ & $-$ \\
$0$ & $1$ & $-2$ & $0p_{3/2}$ & $0.100057870$ & $0.1000005174$ & 1 & $%
0p_{1/2}$ & $0.1232273855$ & $0.1362956215$ \\
$1$ & $1$ & $-2$ & $1p_{3/2}$ & $-$ & $0.1014893336$ & 1 & $1p_{1/2}$ & $%
0.1014382946$ & $0.1707561995$ \\
$2$ & $1$ & $-2$ & $2p_{3/2}$ & $0.100057870$ & $0.1130529719$ & 1 & $%
2p_{1/2}$ & $0.1026480928$ & $0.2164398762$ \\
$3$ & $1$ & $-2$ & $3p_{3/2}$ & $0.123227386$ & $0.1010361207$ & 1 & $%
3p_{1/2}$ & $0.2433208010$ & $0.2731253525$ \\
$0$ & $2$ & $-3$ & $0d_{5/2}$ & $0.123457217$ & $0.1132253472$ & 2 & $%
0d_{3/2}$ & $0.1522417493$ & $0.17104639476$ \\
$1$ & $2$ & $-3$ & $1d_{5/2}$ & $0.1061170006$ & $0.10163993177$ & 2 & $%
1d_{3/2}$ & $0.1923389117$ & $0.2167313327$ \\
$2$ & $2$ & $-3$ & $2d_{5/2}$ & $-$ & $0.10163993177$ & 2 & $2d_{3/2}$ & $%
0.2435547169$ & $0.2734174979$ \\
$3$ & $2$ & $-3$ & $3d_{5/2}$ & $0.1061170006$ & $0.1132253472$ & 2 & $%
3d_{3/2}$ & $0.3056423517$ & $0.3408332607$ \\
$0$ & $3$ & $-4$ & $0f_{7/2}$ & $0.1525865150$ & $0.13658291603$ & 3 & $%
0f_{5/2}$ & $0.1926867981$ & $0.2171368804$ \\
$1$ & $3$ & $-4$ & $1f_{7/2}$ & $0.1000082780$ & $0.1135029417$ & 3 & $%
1f_{5/2}$ & $0.2439041281$ & $0.2738248097$ \\
$2$ & $3$ & $-4$ & $2f_{7/2}$ & $0.0995403787$ & $0.09964377877$ & 3 & $%
2f_{5/2}$ & $0.3059926539$ & $0.34124166206$ \\
$3$ & $3$ & $-4$ & $3f_{7/2}$ & $-$ & $0.09964377877$ & 3 & $3f_{5/2}$ & $%
0.3786566934$ & $0.4190685307$ \\
\tableline\tableline &  &  &  &  &  &  &  &  &
\end{tabular}%
\end{table}


\begin{thebibliography}{99}
\bibitem{1} J. N. Ginocchio, Relativistic harmonic oscillator with spin
symmetry, Phys. Rev. C 69 (2004) 034318-25.

\bibitem{2} J. N. Ginocchio, Pseudospin as a relativistic symmetry, Phys.
Rev. Lett. 78 (3) (1997) 436-439.

\bibitem{3} J.N. Ginocchio, Relativistic symmetries in nuclei and hadrons,
Phys. Rep. 414 (4-5) (2005) 165-261.

\bibitem{4} P.R. Page, T. Goldman, J.N. Ginocchio, Relativistic symmetry
suppresses quark spin-orbit splitting, Phys. Rev. Lett. 86 (2001) 204-207.

\bibitem{5} A. Arima, M. Harvey, K. Shimizu, Pseudo LS coupling and pseudo
SU3 coupling schemes, Phys. Lett. B 30 (1969) 517-522.

\bibitem{6} K.T. Hecht, A. Adler, Generalized seniority for favored $J\neq 0$
pair in mixed configurations, Nucl. Phys. A 137 (1969) 129-143.

\bibitem{7} J.N. Ginocchio, D.G. Madland, Pseudospin symmetry and
relativistic single-nucleon wave functions, Phys. Rev. C 57 (1998) 1167-1173.

\bibitem{8} A. Bohr, I. Hamarnoto, B.R. Motelson, Pseudospin in Rotating
Nuclear Potentials, Phys. Scr. 26 (1982) 267-272.

\bibitem{9} J. Dudek, W. Nazarewicz, Z. Szymanski, G.A. Leander, Abundance
and systematics of nuclear superdeformed states; relation to the pseudospin
and pseudo-SU(3) symmetries, Phys. Rev. Lett. 59 (1987) 1405-1408.

\bibitem{10} D. Troltenier, C. Bahri, J. P. Draayer, Generalized
pseudo-SU(3) model and pairing, Nucl. Phys. A 586 (1995) 53-72.

\bibitem{11} J. Meng, K. Sugawara-Tanabe, S. Yamaji, P. Ring, A. Arima,
Pseudospin symmetry in relativistic mean field theory, Phys. Rev. C 58 (2)
(1998) R628-631.

\bibitem{12} A.D. Alhaidari, H. Bahlouli, A. Al-Hasan, Dirac and
Klein-Gordon equations with equal scalar and vector potentials, Phys. Lett.
A 349 (2006) 87-97.

\bibitem{13} R. Lisboa, M. Malheiro, A.S. De Castro, P. Alberto, M.
Fiolhais, Pseudospin symmetry and the relativistic harmonic oscillator,
Phys. Rev. C 69 (2004) 024319-15.

\bibitem{14} H. Ak\c{c}ay, Dirac equation with scalar and vector quadratic
potentials and Coulomb-like tensor potential, Phys. Lett. A 373 (2009)
616-620.

\bibitem{15} H. Ak\c{c}ay, C. Tezcan, Exact solutions of the Dirac equation
with Harmonic oscillator potential including a Coulomb-like tensor
potential, Int. J. Mod. Phys. C 20 (6) (2009) 931-940.

\bibitem{16} M. Moshinsky, A. Szczepaniak, The Dirac oscillator, J. \ Phys.
A: \ Math. Gen. 22 (1989) L817-L819.

\bibitem{17} V.I. Kukulin, G. Loyola, M. Moshinsky, A Dirac equation with an
oscillator potential and spin-orbit coupling, Phys. Lett. A 158 (1991) 19-22.

\bibitem{18} G. Mao, Effect of tensor couplings in a relativistic Hartree
approach for finite nuclei, Phys. Rev. C 67 (2003) 044318-12.

\bibitem{19} P. Alberto, R. Lisboa, M. Malheiro, A.S. de Castro, Tensor
coupling and pseudospin symmetry in nuclei, Phys. Rev. C 71 (2005) 034313-7.

\bibitem{20} R.F. Furnstahl, J. J. Rusnak, B.D. Serot, The nuclear
spin-orbit force in chiral effective field theories, Nucl. Phys. A 632
(1998) 607-623.

\bibitem{21} M.H. Moshinsky, Y. Smirnov, The Harmonic Oscillator in Modern
Physics, Hardwood Academic Publisher, Amsterdam, (1996) 289-404.

\bibitem{22} M.H. Pacheco, R.R. Landim, C.A.S. Almedia, One-dimensional
Dirac oscillator in a thermal bath Phys. Lett. A 311(2003) 93-96.

\bibitem{23} J.Y. Gou, Z.Q. Sheng, Solution of the Dirac equation for the
Woods-Saxon potential with spin and pseudospin symmetry, Phys. Lett. A 338
(2005) 90-96.

\bibitem{24} A. Arda, R. Sever, Approximate solution of the effective mass
Klein-Gordon equation for the Hulthen potential with any angular momentum,
Int. J. Theor. Phys. 48 (2009) 945-951.

\bibitem{25} Y. Xu, S. He, C.S. Jia, Approximate analytical solutions of the
Dirac equation with the P\"{o}schl-Teller potential, J. Phys. A: Math.
Theor. 41 (2008) 255302-255309.

\bibitem{26} Y. Zhang, Approximate analytical solutions of the Klein-Gordon
equation with scalar and vector Eckart potentials, Phys. Scr. 78 (2008)
015006 (4pp).

\bibitem{27} W.C. Qiang, R.S. Zhou, Y. Gao, Application of the exact
quantization rule to the relativistic solution of the rotational Morse
potential with pseudospin symmetry, J. Phys. A: Math. Theor. 40 (2007)
1677-1685.

\bibitem{28} A. Soylu, O. Bayrak, I. Boztosun, An approximate solution of
Dirac-Hulthen problem with pesudospiun and spin symmetry for any $\kappa $
state, J. Math. Phys. 48 (2007) 082302-9.

\bibitem{29} A. Soylu, O. Bayrak, I. Boztosun, $\kappa $ state solutions of
the Dirac equation for the Eckart potential with pseudospin and spin
symmetry, J. Phys. A: Math. Theor. 41 (2008) 065308 (8pp).

\bibitem{30} C.S. Jia, P. Gao, X.L. Peng, Exact solution of the Dirac-Eckart
problem with spin and pseudospin symmetry, J. Phys. A: Math. Gen. 39 (2006)
7737-7744.

\bibitem{31} L.H. Zhang, X.P. Li, C.S. Jia, Analytical approximation to the
solution of the Dirac equation with the Eckart potential including the
spin-orbit coupling term, Phys. Lett. A 372 (2008) 2201-2207.

\bibitem{32} C.S. Jia, J.Y. Liu, L. He, L. Sun, Pseudospin symmetry in the
relativistic empirical potential as a diatomic molecular model, Phys. Scr.
75 (2007) 388-393.

\bibitem{33} C.S. Jia, P. Gao, Y.F. Diao, L.Z. Yi, X.J. Xie, Solutions of
Dirac equations with the P\"{o}schl-Teller potential, Eur. Phys. J. A 34
(2007) 41-48.

\bibitem{34} H. Motavali, Bound state solutions of the Dirac equation for
the Scarf-type potential using Nikiforov-Uvarov method, Mod. Phys. Lett. A
24 (15) (2009) 1227-1236.

\bibitem{35} C.-S. Jia, T. Chen, L.-G. Cui, Approximate analytical solutions
of the Dirac equation with the quantized P\"{o}schl-Teller potential
including the pseudo-centrifugal term, Phys. Lett. A 373 (18-19) (2009)
1621-1626.

\bibitem{36} X.-C. Zhang, Q.-W. Liu, C.-S. Jia, L.-Z. Wang, Bound states of
the Dirac equation with vector and scalar Scarf-type potentials, Phys. Lett.
A 340 (2005) 59-69.

\bibitem{37} X. Zou, L.-Z. Yi, C.-S. Jia, Bound states of the Dirac equation
with vector and scalar Eckart potentials, Phys. Lett. A 346 (2005) 54-64.

\bibitem{38} G.F. Wei, S.H. Dong, Approximately analytical solutions of the
Manning-Rosen potential with the spin-orbit coupling term and spin symmetry,
Phys. Lett. A 373 (2008) 49-53.

\bibitem{39} K.-E. Thylwe, Amplitude-phase methods for analyzing the radial
Dirac equation: calculation of scattering phase shifts, Phys. Scr. 77 (2008)
065005 (8pp).

\bibitem{40} A.D. Alhaidari, Solution of the Dirac equation by separation of
variables in spherical coordinates for a large class of non-central
electromagnetic potentials, Ann. Phys. 320 (2005) 453-467.

\bibitem{41} C. Berkdemir, R. Sever, Pseudospin symmetry solution of the
Dirac equation with an angle-dependent potential, J. Phys. A: Math. Theor.
41 (2008) 045302 (11pp).

\bibitem{42} S.M. Ikhdair, R. Sever, Approximate analytical solutions of the
general Woods-saxon potentials including the spin-orbit coupling term and
spin symmetry, to appear in Cent. Eur. J. Phys. (2010).

\bibitem{43} S.M. Ikhdair, Approximate solutions of the Dirac equation for
the Rosen-Morse potential including the spin-orbit centrifugal term, to
appear in J. Math. Phys. 51 (2) (2010).

\bibitem{44} S.M. Ikhdair, An improved approximation scheme for the
centrifugal term and the Hulth\'{e}n potential, Eur. Phys. J. A 39 (2009)
307-314.

\bibitem{45} S.M. Ikhdair, R. Sever, Any $l$-state improved quasi-exact
analytical solutions of the spatially dependent mass Klein-Gordon equation
for the scalar and vector Hulth\'{e}n potentials, Phys. Scr. 79 (2009)
035002 (8 pages).

\bibitem{46} C.S. Lam, Y.P. Varshni, Energies of s eigenstates in a static
screened Coulomb potential, Phys. Rev. A 4 (1971) 1875-1881.

\bibitem{47} B. Durand, L. Durand, Duality for heavy quark systems, Phys.
Rev. D 23 (1981) 1092-1102.

\bibitem{48} R.L. Hall, Envelope representations for screened Coulomb
potentials, Phys. Rev. A 32 (1985) 14-18.

\bibitem{49} R. Dutt, K. Choudhury, Y.P. Varshni, An improved calculation
for screened Coulomb potentials in Rayleigh-Schrodinger perturbation theory,
J. Phys. A: Math. Gen. 18 (1985) 1379-1388.

\bibitem{50} J. Lindhard, P.G. Hansen, Atomic Effects in Low-Energy Beta
Decay: The Case of Tritium, Phys. Rev. Lett. 57 (1986) 965-967.

\bibitem{51} I.S. Bitensky, V.K. Ferleger, I.A. Wojciechowski, Distortion of
H2 potentials by embedding into an electron gas at molecule scattering by a
metal surface, Nucl. Instrum. Meth. B 125 (1997) 201-206.

\bibitem{52} C.-S. Jia, J.Y. Wang, S. He, L.-T.Sun, Shape invariance and the
supersymmetry WKB approximation for a diatomic molecule, J. Phys. A: Math.
Gen. 33 (2000) 6993-6998.

\bibitem{53} J.A. Olson, D.A. Micha, Transition operators for atomatom
potentials: The Hilbert Schmidt expansion, J. Chem. Phys. 68 (1978)
4352-4356.

\bibitem{54} A.F. Nikiforov, V.B. Uvarov, Special Functions of Mathematical
Physics, Birkhauser, Bassel, 1988.

\bibitem{55} S.M. Ikhdair, R. Sever, Improved analytical approximation to
arbitrary $l$-state solutions of the Schr\"{o}dinger equation for the
hyperbolical potentials, Ann. Phys. (Berlin) 18 (10-11) (2009) 747-758.

\bibitem{56} S.M. Ikhdair, Exact Klein-Gordon equation with spatially
dependent masses for unequal scalar-vector Coulomb-like potentials, Eur.
Phys. J. A 40 (2) (2009) 143-149.

\bibitem{57} J.D. Bjorken, S.D. Drell, Relativistic Quantum Mechanics,
McGraw-Hill, NY, 1964.

\bibitem{58} G.R. Satchler, Direct Nuclear Reactions, Oxford University
Press, London, 1983.

\bibitem{59} Z.-Y. Chen, M. Li , C.-S. Jia, Approximate analytical solutions
of the Schr\"{o}dinger equation with the Manning-Rosen potential model, Mod.
Phys. Lett. A 24 (23) (2009) 1863-1874.

\bibitem{60} Y.-F. Diao, L.-Z. Yi, T. Chen, C.-S. Jia, Arbitrary $l$-wave
bound state solutions of the Schr\"{o}dinger equation with the Eckart
potential, Mod. Phys. Lett. B 23 (2009) 2269-2279.

\bibitem{61} T. Chen, J.-Y. Liu, C.-S. Jia, Approximate analytical solutions
of the Dirac-Manning-Rosen problem with the spin-symmetry and pseudo-spin
symmetry, Phys. Scr. 79 (2009) 055002 (7pp).

\bibitem{62} T. Chen, Y.-F. Diao, C.-S. Jia, Bound state solutions of the
Klein-Gordon equation with the generalized P\"{o}schl-Teller potential,
Phys. Scr. 79 (2009) 065014 (6pp).

\bibitem{63} S.M. Ikhdair, Rotation and vibration of diatomic molecule in
the spatially-dependent mass Schr\"{o}dinger equation with $q$-deformed
Morse potential, Chem. Phys. 361 (1-2) (2009) 9-17.

\bibitem{64} S.M. Ikhdair, Bound states of the Klein-Gordon equation for
vector and scalar general Hulth\'{e}n-type potentials in $D$-dimension, Int.
J. Mod. Phys. C 20 (1) (2009) 25-45.

\bibitem{65} W.-C. Qiang, R.-S. Zhou, Y. Gao, Any $l$-state solutions of the
Klein-Gordon equation with the generalized Hulth\'{e}n potential, Phys.
Lett. A 371 (2007) 201-204.

\bibitem{66} F. Benamira, L. Guechi, A. Zouache, Comment on 'Exact solutions
of the s-wave Schrdinger equation with Manning-Rosen potential', Phys. Scr.
80 (2009) 017001-3.

\bibitem{67} I.S. Gradshtein, I.M. Ryzhik, Tables of Integrals, Series and
Products, Academic, New York, 1969.
\end{thebibliography}
\end{document}